# PROSPECTS AND CHALLENGES IN DIGITIZATION: THE CASE STUDY OF FEDERAL PARLIAMENT OF NEPAL

**Arun Kishor Sharma[1] & Dr. Sandeep Kautish[2]**
[1] M. Sc. (ITM) Scholar, Lord Buddha Education Foundation, Kathmandu, Nepal
[2] Dean (Academics), Lord Buddha Education Foundation, Kathmandu, Nepal

### Abstract

Digitization means the use of digital technology to modify a business model to create new possibilities for sales to value creation. The main purpose of the study is to identify the effects of digitization in the federal parliament of Nepal. The performance of the parliament, secretariat, parliamentarians as well as employees working procedure should be fully digitalized based different prescribed different modules. Quantitative and quantitative method were followed. This research included a series of well-structured questionnaires, mainly directed at MPs and employees, and structured interviews with key persons of the organizations. The raw data were processed and analyzed by SPSS. Cochran's Q test, Chi-Square test and Friedman tests were applied to show the present status and need of digitization tools in Federal Parliament of Nepal. The system becomes automated, interlinked, based on the database and there should be a mutual understanding among committees on different issues. The overall activities of parliament, committees, and its secretariats records are managed systematically through tailored software. We would also intend to discuss the limits on the usage of different tools and recommendations for parliamentary application. The holistic framework is a digital framework for FPN. This will be helpful for a digitized workplace. The FPN will start the new era of digital transformation.

**Keywords** - Federal Parliament, Digitalization, Secretariat, Parliamentarian, Legislature, Information, and Communication Technology

## 1 INTRODUCTION

Usually, parliamentarian visits worldwide are a result of an invitation to share best practices and address common challenges among representative legislatures in the Westminster system. There are numerous common themes, including the legislative process (bills becoming laws), and the large quantity of physical paperwork generated in printing and reviewing draft legislation in paper form. After sharing and discussing among the parliamentarian of well-developed countries, our parliamentarians also acknowledged the need of aligning the Nepali parliamentary processes with available technology (Kaur and Kautish, 2019, 2018). Accordingly, they have to introduce the use of Information and Communications Technology (ICT) in their Strategic. Digitization of Federal Parliament of Nepal (FPS) to enhance the core functions of parliament by improving effectiveness, accountability, transparency, and inclusivity using ICT (Liringstone & Smith, 2014).

### 1.1 Introduction to Federal Parliament

The Federal Parliament has two chambers of parliament. There are 275 members elected by the House of Representatives (HR) for a five-year term, 165 from single-seat constituencies and 110 from a proportional party list. Similarly, there is a system of the National Assembly (NA) of 59 members elected for a six-year term from 7 provinces, equivalent to 8 members. The President nominates the remaining three representatives, including one person, 1 Dalit, and 1 from the community of people of different capacities. (Secretariat, 2020)

Both Houses (HA and NA) of FPS have a strong secretariat under the Speaker and Chairman to maintain the effectiveness and efficiency of parliamentary business. Administration of both the assemblies (Lower and Upper house) is managed under the leadership of the Chief Secretary of the Parliament under the direct supervision and guidance of Rt. Honorable Speaker of HA and Rt. Honorable Chairperson of the NA.

## 1.2 Ethical issues, challenges

Verbal informed consent was taken from each respondent before data collection. Ethical approval was obtained from the University Ethics Committee (UEC) of Asia Pacific University. International Research Committee (IRC), is the ethical consideration for my research study.

## 1.3 Problem Statement

The documentations in FPS and its secretariat are the property of nations. They are a fully manual-based filing system'. They are not well arranged. We can't find the documents, decisions, and reports. They are based on institutional memory. The decisions are taken in FPS and its committees are property of states. So, full automation i.e., a digitization system is required.

The dissertation shows the various aspects of digitization in the federal parliament of Nepal.

- Focusing on a common platform where the citizen can and policymaker can communicate via the popular projects like LEX-IS, +Spaces, NOMAD, ARCOMEM, METALOGUE used worldwide in the world. (Fotios Fitsilis, 2017)
- Introduction of www.lipad.ca, an online portal built as a repository for Canadian government data archiving, with legislative hearings at the core of its architecture. (Kaspar Beelen, 2017)
- Interactive, participatory, and advisory forum for citizens' participation in the decision-making process through the use of ICT mechanisms to enhance parliamentary procedures, services, and functions. (Abdulsalam S. Mustafa, 2018)
- Citizen's participation in law-making using smartphones and computers. "e-Tegeko" is a kind of the proposed solution for digitization. (Rene Kabalisaa, 2016)

Below are the main issues which are not still focused on by researchers so far.

- Mostly the papers were focused on some of the projects like LEX-IS, +Spaces, NOMAD, ARCOMEM, METALOGUE, and e-Tegeko, etc. Hence, to determine the actual implementation of digitalization, the different modules need to be focused on.
- No projects were found in the case of the Federal Parliament of Nepal. Hence, it would not be appropriate to use those projects prescribed by the researchers in the context of Nepal as compared to other countries' parliaments.

## 1.4 Research Questions

- RQ1 What is the current status of digitalization in the FPN?
- RQ2 What are the challenges and barriers to implementing the digitization of the FPN?
- RQ3 What are the pre-requisites that can be suggested to regulators for smooth implementation in the digitization of the FPN?

## 1.5 Aims

The primary objective of the test analysis would be to classify the results of digitization in the FPS. It will determine the path to get the result of the whole research study. The aim of the research is to how digitization will bring an opportunity to parliament and the secretariat. The new idea will find out from the research study.

## 1.6 Objectives

The main objective is to find the benefits and advantages of using digitization in the FPN.

- Specific Objective
- To determine the current status of digitization in the FPN.
- To evaluate the challenges and barriers to implementing the digitization of the FPN.
- To examine the pre-requisites to regulators for smooth implementation in the digitization of the FPN.

## 1.7 Scope and significance of research

FPS and its secretariat are the main organs of the state. The scope of this research is to implement the use of digitization in the FPS. So, the research should be conducted within the FPS and its secretariat. This research paper is beneficial both for FPS and its administrative body Federal Parliament Secretariat (FPS) for future purposes.

This research should be a great contribution to the FPN and its FPS. The paper should cover the following features which are mentioned below:

- The legislature's work should be digitalized rather than based on a manual system.
- The MPs and employees can get the document on time so that their decision power should be on time.
- The exchange of the documents within section, department, and committee is easier.
- The chances of dispute within the committee are lesser.
- This system helps in making better laws and policies in comparison with the previous and existing system.

## 1.8 Limitations of the study

Though there are many challenges for the digitization of parliamentary processes, some of them are explained below:

- Lack of skilled human resources in terms of skills as well as numbers. so, instruction reforms are needed.
- Lack of budget: FPN has to depend on the Ministry of Finance (MOF).
- Lack of infrastructure: The information technology department has been limited to a single room.
- Constraints to the success may be e-illiteracy; lack of skills; limited infrastructure; failure to pass new laws; Internal resistance to transformation etc.

## 2 LITERATURE REVIEW

The section shows reviews of several works of literature by clustering them on the justification of the main points such as digitization, benefits of digitization, parliament case studies of different countries.

Digitization is a method in which an analog signal is captured in digital form (Stevan Gostojic, 2020). It is the process of creating an electronic means of a 'real world' object or case, allowing the object to be stored, viewed, and processed on a computer, a 2018th shared over networks and/or the World Wide Web. The word 'digitization' is a simplified expression of the document in digital format (Julia Schwanholz, 2018). The following are some of the benefits of digitization.

- Access: Digitized document offers a retrieval benefit over the physical document (Schumacher, 2016).
- Preservation: It is possible to copy digital content and it does not rely on having a constant item and being held under guard, but on the potential to deliver several copies (Darko Cherepnalkoski, 2015).
- Reduced costs of Handling: The cost of processing, preserving, and copying paper records is minimized by digitization, and the missing documents can be restored in some cases (Beland & Muphy, 2016).
- Organization and dissemination: You can organize digital or electronic objects and archive them in a file data warehouse.

**Indian Experience**

The website is updated regularly with information about the acts of Parliament and legislative news. The Blog, Twitter, and Facebook pages make it easy for people to get involved. The journalists attend workshops on monitoring the actions of MPs and MLAs. It also offers information on the legislative agenda in Parliament, as well as details on legislative results, to the press and electronic media. Members of the PRS staff are often asked to contribute columns on a variety of important bills (PRS Legislative Research, 2020).

The Parliament of India initiated a parliamentary digitization project in 2012, with an initial focus on digitizing parliamentary debates and records since 1858.  A web portal was also developed for the dissemination of information for public access to legislation and documentation. The available documentation includes Lok Sabha debates from the first to the 11th Lok Sabha (Parliament) covering 45 years (1952-1997) and reports of all parliamentary committees from 1952 to 1998.   The third phase of the project involved the digitization of the relevant publications of the Lok Sabha

Secretariat and historical debates of — Provisional Parliament (1950-1952); Constituent Assembly of India (Legislative) (1947-1949), Central Legislative Assembly (1921-1947); Council of State (1921-1954) and Indian Legislative Council (1858-1920) (Rajya Sabha, 2019).

The public web portal contains numerous links of both houses of Parliament of India including MP bios, a daily parliamentary calendar, pending legislation, bills, committee business, real-time (live streaming), and house and committee debates.   The Right to Information Act (2005) is available in full text and synopsis form and provides a specific reference to the scope of external (citizen, NGO, other stakeholders) and internal (parliamentary) response and binding information release to interested parties (Indian Emblem, 2020).

Media access is facilitated with on-line access to the parliament's Press Information Bureau and comprehensive live streaming of parliamentary proceedings is also available, in addition to a general information page rich with links to information regarding the legislative process and the structure and function of the people's house—with specific reference to Parliament's legislative, oversight, and representative functions whilst highlighting Parliament's mandate to maintain its status as an open, transparent and accountable institution (LSTV, 2014).

**EU Experience**

Since 1995, technical convergence has blurred the distinctions between telecommunications, television, and information technology. In 2015, the Commission initiated the digital way to deliver the key policy proposals, including strengthening e-commerce, copyright, e-Privacy, digital rights harmonization, and cybersecurity. (Digital Agenda for Europe, 2020).

There are multiple channels available to communicate directly with MPs within the European Parliament, including the Internet, social networks, and video broadcasts of almost all meetings in full session, on a committee-by-committee basis, and on personal MP's social media sites—all of which are released publicly via the European Parliament website. (Stay informed, 2020).

Likewise, it has become customary for MEPs (and staff) to make themselves highly accessible to traditional media (television, radio, newspapers) while at the same time promoting individual, country-specific, and broader EU themes interactively with very large and diverse participants through the Union (Digital Agenda for Europe, 2020).

Transparency, accountability, and inclusivity are evident in nearly all entry points for external stakeholder engagement—although there are increasing instances of division and competing interests between national (e.g., member state) and unified EU policy and procedure, particularly when comparing the relative level of broadband penetration, infrastructure development, and familiarity amongst the individual member state populations with ICT.  This gap in capacity and knowledge/familiarity is narrowing, however, as newer member states "catch up" to the more established and some extent more developed member states (Essential ICT infrastructure for smart, automated transport, 2020).

The Parliament also provides IT support and advice for MPs in both houses, including drop-in IT support at MP's request (with such requests made via various media and web-based interactive software).  The PDS also manages the parliamentary IT network and ensures that internal computer systems and equipment are running properly and are provided software updates on a regular and periodic basis.  The department is also charged with developing specialized applications for the evolving needs of Parliaments whilst developing an ICT strategic plan that is periodically reviewed, updated, and amended to help ensure cost-effectiveness, human resource relevance, and usefulness of all ICT systems to internal (MPs, staff) and external (the public at large) stakeholders.

Finally, PDS maintains the Parliament website, parliamentary social media accounts (Twitter, Facebook, Instagram), and the intranet.  There are also existing guidelines for Parliament on the proper use of social media platforms in the parliamentary context (Essential ICT infrastructure for smart, automated transport, 2020).

# 3  RESEARCH DESIGN AND METHODOLOGY

## 3.1 Introduction

This part aims to use mixed study techniques such as surveys, questionnaires, and case studies to formulate the research methodology. This approach to research tries to help experts found the outcome. The chapter contains the design of the research methods which illustrates the procedure of performing the research during the study period. In this part, the methods used for data collection are also explained. Here, the researcher identified the research design accompanied by the methods of qualitative and quantitative data collection, ethical issues of research, and the shortcomings of the field of analysis.

### 3.2 Research Strategy

An overall plan for the execution of a research report is a research strategy. In preparation, implementation, and tracking the analysis, a research approach directs a researcher. Although the research strategy offers valuable support to a greater extent, a research methodology that can direct research work at a more comprehensive level need to be accompanied. For its intent, a study technique should be acceptable, i.e., it should be able to assist the researcher to find a response to the research issue being considered (Geoffrey R. Marczyk, 2005).

### 3.3 Research Method

Research Methodology refers to the multiple procedural actions that a researcher would take in researching a topic with such targets in mind. This explains the mechanism and procedure used in the entire topic of the analysis. That is the way to consistently deal with the issue of science. It is the framework of the proposal and analysis approach intended to address the research questions or validate the research hypothesis. The nature of the study is to regulate variance. This covers multiple dependent and independent variables, study design styles, test questions, survey theories, data collection tasks, analytical methods, etc. Also, the comprehensive plan for the study of science is the technique. Different forms of research methods, i.e. qualitative, quantitative, and mixed methods, have been used in academic fields. etc. (Igwenagu, 2016). This thesis focuses on the study of ICT in the parliament of Nepal. Research data was collected from primary data through the parliamentarian and employee. The questionnaires are requested through the online method due to COVID-19. The final results of the research project are aimed at the effective implementation of ICT in the parliament of Nepal. The above modes of research for efficient and effective approaches to Digitization.

### 3.4 Research Approach

A research approach is a technique and procedure composed of the general principles of systematic methods of data collection, measurement, and analysis (Khanal, 2020). It is applied based on empirical issues that need to be discussed.  The approach is categorized in the following ways:

    i.   The approach of data collection
   ii.   The approach of data analysis

*Figure 1: Components of Research Approach*

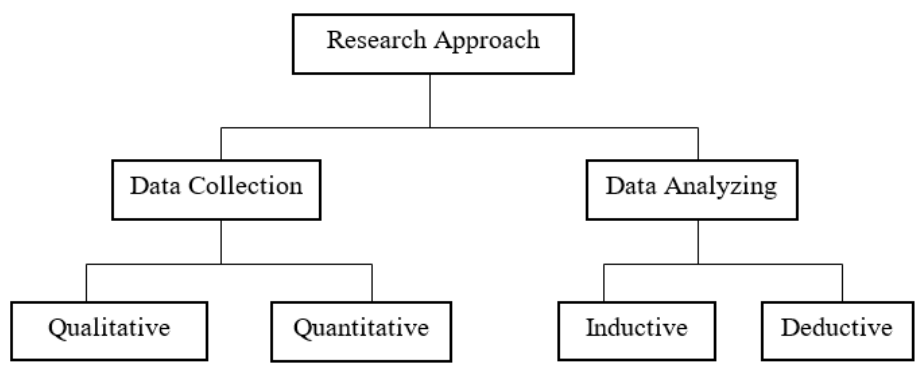

### 3.5 Data collection method and tools for data collection

The primary data was obtained exclusively from Microsoft forms in an electronic medium. Microsoft forms are available online to allow people to respond to them. The questionnaires were based on numerous variables that were found in other literature studies. The questionnaire contained standard questions such as the age group, the availability of administrative materials, the use of various technologies and applications, the expectations of different approaches by the respondents. The data obtained was then shipped to Microsoft Excel and the data processing in SPSS software. Some of the tools for data collection were questionnaire, telephone, mobile phone and facsimile, mail, and interview.

### 3.6 Sample Selection

The sampling method for my research was the convince sampling model i.e., also called availability sampling. It was the primary source of data collection where data was collected from available members. In other words, we can say that data was collected only with their consent.

### 3.7 Analysis Plan

Research data analysis is a process used by researchers for reducing data to a story and interpreting it to derive insights. The data analysis process helps in reducing a large chunk of data into smaller fragments, which makes sense. They are qualitative and quantitative data analysis.

**Quantitative analysis**

Quantitative analysis uses numerical data to identify statistical relationships between variables. Quantitative data are numerical, ordinal, nominal. For example, surveys, questionnaires, and evaluations that include multiple-choice items and ratings (e.g., Likert scale) provide quantitative data for analysis (Geoffrey R. Marczyk, 2005).

**Qualitative analysis**

Qualitative research uses insightful evidence to explain processes (e.g., how students interact in a group), gain perspectives into the type of principles that increase knowledge, and present the world view from the participants' point of view (e.g., the teachers, students, and others related to the classroom). Qualitative data are descriptive. For example, field notes, interviews, video, audio, open-ended survey questions all provide qualitative data for analysis (Igwenagu, 2016).

For the research topic on "Prospects and Challenges in Digitization: The case study of Federal Parliament of Nepal", quantitative research and quantitative method were followed. In other words, we can say that we follow the mixed approach. This research included a series of well-structured questionnaires, mainly directed at MPs and employees, and structured interviews with key persons of the organizations.

## 4   DATA ANALYSIS AND FINDINGS OF RESEARCH

### 4.1 Reliability test analysis (Cronbach's Alpha Reliability Test)

This test was done to test the reliability. The questionnaire is distributed to 40 people to FPN. The result was as follows:

*Table 1: Result of Pilot Study*

| Reliability Statistics | |
|---|---|
| Cronbach's Alpha | N of Items |
| .704 | 5 |

The value of Cronbach's Alpha is more than 0.70, then the research is termed as good.

To meet the expectation of the research, a survey was done in FPS among 137 Member of parliament and employees of FPS. There were 137 respondents. Among them, 19 were members of parliament and 118 were an employee of a different post. The age group between 20-29, 30-39, 40-49, and 60 above had taken part in the survey. The age group between 30-39 had 60 respondents whereas the age group between 60 and above had only one respondent. There were altogether 58.39% male and 41.61% were female. The study is more precise and accurate because all categories of respondents have been taken in the study area. Out of the total 137 respondents, 13.87% were members of Parliament, 5.11% were Joint Secretary, 10.95% were Under secretary, 24.82% were officers and 45.26% were non-gazette officers.

# 5 Discussion

## 5.1 Summary

This study is very vital for FPS because this is the first time a study was done in the context of the FPS. The first research question stated the current status of digitization in the FPS whereas the second question focused on challenges and barriers to implementing digitization. The last question emphasized on pre-requisite for the smooth implementation of digitization

## 5.2 Research questions and findings of the survey

According to the first research questions, we concluded that among different means of digitization such as website, mobile APPS, and intranet. The website became the most available digitization tool in the federal parliament of Nepal. In the same way, another test called Chi-Square had been applied to find the access of the digitization tools in terms of job position, age group, and gender. The result based on job position indicated that post and mode of access of information by using digitizing tools had no relations. In the same way, gender also did not matter for accessing information whereas age does not matter. In the FPS, the two digital devices webpage and APP were available tools. By applying the chi-square test, we found the frequency of uses of Webpage and App based on the job position, age groups, and gender. The usage of the frequency of digitization tools was not associated with job positions and age whereas gender was associated with accessing the information.

According to the second research question, we had identified some of the barriers and challenges like Cyber Security, lack of IT knowledge, and lack of digital authentication process. The above-mentioned elements were equally challenging or at least one of the modes was challenging could be found by using the Friedman test for barriers/ challenges for digitization. The result showed the IT knowledge was the most challenging barrier in digitization. We had also analyzed the perceptions of employees and members of parliament in terms of digital transformation in the organization. The Chi-square test had been applied on the following variables as Job position/ age groups/ genders. The results proved that lack of expertise, limited budget, lack of technology, employee resistance, and not known were equally challenging based on position and gender whereas age did not matter.

The last research questions prerequisites for implementing digitization could be done by using the Friedman test. The result proved that Management Information System (MIS) and Training were equally important. In another part of the test, we had compared Gate Pass System (GPS), Committee and Meeting Management System (CMMS), and Assembly Questions Processing System (AQPS). From the above test, it proved that all the modules were equally important for the Federal Parliament of Nepal (FPS).

## 5.3 Similarities of Literature Review

The journal " Implementing Digital Parliament Innovative Concepts for Citizens and Policy Makers" Focuses on a common platform where citizen and policymaker could communicate via the popular projects like LEX-IS, +Spaces, NOMAD, ARCOMEM, METALOGUE used worldwide in the world. (Fotios Fitsilis, 2017). In the same way, the Canadian parliament also highlighted the web portal called www.lipad.ca, an online portal built as a repository for Canadian government data archiving, with legislative hearings at the core of its architecture. (Kaspar Beelen, 2017).

The journal named "The Challenges of e-Parliament Adoption and its Mitigation" focused on the interactive, participatory and advisory forum for citizens' participation in the decision-making process through the use of ICT mechanisms to enhance parliamentary procedures, services, and functions. (Abdulsalam S. Mustafa, 2018). "e-Tegeko" is a kind of the proposed solution for the digitalization in the parliament of Rwanda (Rene Kabalisaa, 2016 ).

The report introduced the possibilities resulting from the use of ICT in Parliament and its development. This suggested that frameworks such as e-Parliament or Virtual Parliament where people and politicians work together, coordinate, or merely connect, with a shared interface. Parliament is an institution for democracy. So, we could not rely on the traditional way. So, they focused on some of the projects used in some countries of parliament. They were as LEX-IS, +Spaces, NOMAD, ARCOMEM, METALOGUE. This paper only discussed some of the projects used worldwide in the world (Fotios Fitsilis, 2017).

It discusses the scope of the enforcement of ICT in the legislature. The Legislature offers an open, participatory, and consultative forum for citizens' insight into the law-making and decision-making process through the use of ICT technologies to enhance legislative procedures and resources in any country. In dealing with security concerns, the major problems examined included minimal tools accessible to parliaments, inadequate technical expertise among legislative personnel, low citizen participation, and lack of uniform software.  The Solutions suggested to include the implementation of an e- parliament's s-by the use of data science and the IoT, and storage data, the use of structured XML-based applications, enhanced ICT security in the decision-making process (Abdulsalam S. Mustafa, The Challenges of e-Parliament Adoption and its Mitigation, June 2018).

### 5.4 Validation of the proposed framework

To check its effectiveness of framework, the consultation with Mr. Suraj Dura, a Joint-sectary, Mr Tribikram Parajuli Under-Secretary of FPS. They found the purposed framework will be a holistic approach for a digital transformation for FPS. They commented the purposed framework is a good solution for creating digital transformation in FPN.

## 6  Conclusion and Recommendation

### 6.1 Conclusion

Digitization is one of the basic needs in the 21st century but there is a gap between the available technology and the daily 'real-time' usage of digitalized information in the Nepali parliament at the federal and provincial levels.

There are some specific areas for further identification, assessment, and follow-on implementation identified in this research and those which are specific to the various components of ICT implementation in a parliamentary context.  These include the hardware and software which would need to be procured following Nepal government regulations and as part of any potential funding arrangement between the government and donor agencies.

Suffice it to say, this research is intended to be a framework for digitization and its relevant parliamentary stakeholders might together construct a strategic plan for ICT implementation at both levels of parliament moving forward.  The primary elements of any strategic plan (people, money, and time) must be considered in that formulation to first identify all possible implementation strategies and then to identify which priorities for ITC enhancement are feasible given available human and financial resources within a short-, medium-, and long-term time frames.

As evidenced from the research the preliminary assessment of the ITC needs of Parliament at the federal, there is a gap between the available technology and the daily 'real-time' usage of digitalized information in the Nepali parliament.

There are several specific areas for further identification, assessment, and follow-on implementation identified in this article template and those which are specific to the various components of ICT implementation in a parliamentary context.  These include the hardware and software which would need to be procured following Nepal government regulations and as part of any potential funding arrangement between the government and federal parliament.

### 6.2 Recommendation

Based on the above finding and research the following targets will be set to achieve the research objectives:

Target 1: To Automate Parliament Administration

Target 2: To Implement Legislative Support Systems

Target 3: To Make System Sustainable & Continuous enhancement of the digitized system.

There are altogether three targets for the implementation of digitization. Each target will have a strategy. They are as follow:

Target 1: To Automate Parliament Administration

Strategy 1.1: Need for Infrastructure Development

*Table 2: Need of infrastructure in FPS*

| Activities | Implementation Indicators |
|---|---|
| Preparation of Data Center (Temporary DC) | Data Center (DC) is created. |
| Formulation of Policies relating to ICT Infrastructure | Needed ICT policy |
| Development of Detail Project Report for ICT Infrastructure of Parliament | Needed |
| tallation & Improvement of Network within Parliaments | GB network established in Parliaments |
| Creation and Implementation of Data Center (DC) and Disaster Recovery Center (DRC) with High Availability (HA) Configuration | Data Center (DC) and Disaster Recovery Center (DRC) are created. |

Strategy 1.2: Development and Implementation of Software Modules to be used to Automate Parliament Administration
- Management Information System (MIS)
- Gate Pass System (GPS)
- Assembly Question Processing System (AQPS)
- Committee and Meeting Management System (CMMS)
- Financial Management System (FMS)

Strategy 1.3: Implement Media Technology
- Audio and Visual Recording System with Repository
- Live Broadcasting via social media & mobile app
- Conferencing with Local & Province Government
- Establishment of Media Center
- Implementation of Digital Signage

Strategy 1.4: Capacity Building and Information Exchange with Stakeholders
- Training Sessions
- Conduct Information Exchange Programs with stakeholders

Target 2: To Implement Legislative Support System
Strategy 2.1: Implement Groupware System
- Implement Secured Groupware System to enable parliament users and officials to communicate between them in a secured manner.

Strategy 2.2: Implementation of Knowledgebase and Portal Services
- Development and Implementation of Legislative Support System
- Implementation of Right to Information System

Target 3: To Make System Sustainable & Continuous Enhancement
Strategy 3.1: Develop ICT and Media Directives
Develop ICT Policy and Directive and implement for organized ICT development in Parliament that include but not limited to;
- ICT Enabled Parliament Regulation
- Hardware Replacement Policy
- Hardware Up-gradation Policy
- Software Development Policy
- Software Enhancement Policy
- ICT Security Policy
- Media Rules and Filming Regulations
- Content Broadcasting Policy etc.

Strategy 3.2: O & M Survey and management of ICT Organization
- Organization and Management (O&M) Survey for Strong ICT Organization of Parliament
- Establishment of Training Center

Strategy 3.3: Conduct Training Sessions for MPs, Administers and officials of Parliaments

- Computer and ICT Training to MPs, their assistants, and Officials